\documentclass[12pt]{iopart}
\usepackage{iopams}
\usepackage[dvips]{graphicx}
\usepackage{epsfig}
\usepackage{dcolumn}
\usepackage{bm}

\begin{document}
\title{Temperature dependent electronic correlation effects in GdN}

\author{A.Sharma and W.Nolting}
\address{Institut f$\ddot{u}$r Physik, Humboldt-Universit$\ddot{a}$t zu Berlin, Newtonstr.15, 12489, Berlin, Germany.}
\ead{anand@physik.hu-berlin.de}

\begin{abstract}
We investigate temperature dependent electronic correlation effects in the conduction bands of Gadolinium Nitride (GdN) based on the combination of many body analysis of the multi-band Kondo lattice model and the first principles TB-LMTO bandstructure calculations. The physical properties like the quasi-particle density of states (Q-DOS), spectral density (SD) and quasi-particle bandstructure (Q-BS) are calculated and discussed. The results can be compared with spin and angle resolved inverse photoemission spectroscopy (ARIPS) of the conduction bands of GdN. A redshift of 0.34 eV of the lower band edge (T=$T_{c}$ $\rightarrow$ T=0) is obtained and found in close comparison with earlier theoretical prediction and experimental value reported in the literature.
\end{abstract}


\maketitle

\section{\label{sec:intro} INTRODUCTION}

\indent The rare earth monopnictides and monochalcogenides form a very interesting group of materials. They exhibit a rich variety of anomalous physical properties attracting a considerable attention for basic and applied research. Among the rare earth alloys, the isomorphic compounds like $Gd^{3+} X^{3-}_{p}$, where $X_{p}$ = N,P,As,Sb,Bi and $Eu^{2+} X^{2-}_{c}$, where $X_{c}$ = O,S,Se,Te can be compared since they are isoelectronic in nature. If pure ionic bonding is considered then they are expected to be insulators or semiconductors. The divalent rare earth monochalcogenides are indeed insulators or semiconductors\cite{wachter79,mauger} but the trivalent rare earth monopnictides, especially the rare earth nitride GdN, show an intricate conducting character. For instance, the bandstructure calculations suggest GdN to be a semiconductor\cite{hasegawa,lambrecht,ghosh},  half-metal\cite{aerts,duan,eyert} or a semiconductor in the paramagnetic and a semimetal in the ferromagnetic case using quasi-particle self energy corrections\cite{petukhov}. The above mentioned results used different computational techniques. Hasegawa et.al,~\cite{hasegawa} were the first to calculate the self-consistent electronic bandstructures of GdN for the paramagnetic case using augmented plane wave (APW) method. Their calculations employed one-electron potential instead of local spin density (LSD) functional theory. While Lambrecht\cite{lambrecht} addressed the problem by estimating the gap corrections beyond local density approximation (LDA), Ghosh\cite{ghosh} performed self consistent spin polarized calculations using full potential linear muffin-tin orbital (FP LMTO) but with rigid shifts in the 5{\it d} and 4{\it f} states in order to fit the experimental X-ray photoemission spectroscopy (XPS) and X-ray Bremsstrahlung Isochromat spectroscopy (BIS) results. The Self Interaction Corrected local spin density (SIC-LSD) calculations by Aerts et.al,~\cite{aerts} and augmented spherical wave (ASW) within LDA and generalized gradient approximations (GGA) by Eyert\cite{eyert} rendered half-metallic nature to GdN. There has also been an interesting investigation on the electronic structure and magnetic properties of GdN by Duan et.al,~\cite{duan} based on first principles calculations as a function of unit cell volume. They found that GdN transforms first from half-metallic to semi-metallic and then finally to a semiconductor upon applying stress. These features reveal a strong lattice constant dependence on the electronic structure of GdN. However, experimental results demonstrated bulk GdN to be a low carrier semimetallic\cite{wachter80} or GdN thin films to be insulating\cite{xiao}. \\
\indent The magnetic properties of GdN also form a very interesting study and the mechanism of magnetism is highly intriguing. There had also been a dispute regarding the magnetic properties~\cite{wachter80} of GdN with earlier reports describing it to be an antiferromagnetic\cite{wachter80,cutler} material while other studies indicated it to be a ferromagnet\cite{boyd,gambino}. But after such controversial discourse, it has been accepted that stoichiometric GdN is a ferromagnetic\cite{li,leuenberger} material. There have been theoretical attempts in order to explain the magnetism in GdN using RKKY interaction picture by Kuznietz\cite{kuznietz} and a model proposed by Kasuya\cite{kasuya} which is related to the cross processes between the f-d mixing and the f-d exchange interaction. In this paper we do not discuss the mechanism of such anomalous behaviour of magnetic properties in GdN but it forms a part of future investigations. As a whole, it generates a huge motivation to study such an interesting material and understand the physical phenomenon in similar strongly correlated systems.\\
\indent Here we combine the many body analysis of the multi-band Kondo lattice model alongwith the first principles TB-LMTO spin-polarized bandstructure calculations to investigate the temperature dependent electronic correlation effects in the conduction bands of GdN. We wish to emphasize that upon using bandstructure calculations we are not aiming to find the nature of the electronic ground state of GdN but rather take its output as the starting point (input) for our many body theory and compare the results for the temperature dependence with that of the experimental ones (redshift phenomenon). Very recently a similar study has been reported\cite{bhattacharjee} but using a different approach for determining the self-energy\cite{nolting96} and also for the input in the many body part. \\
\indent The layout of the paper is as follows. In section~\ref{sec:th-mo-cal} we develop the complete multi-band Hamiltonian. A self energy ansatz, authentic for low carrier densities, is used to solve the model Hamiltonian. In section~\ref{sec:bndstr-GdN}, we present electronic bandstructure of the conduction bands of GdN using tight binding linear muffin-tin orbital (TB LMTO) program within LSDA and also obtain the inter-band exchange coupling. In the next section~\ref{sec:recal-GdN}, the many body model is combined with first principles T=0 bandstructure calculations in order to calculate the temperature dependence of some physical properties like quasi-particle density of states (Q-DOS), spectral density (SD), quasi-particle bandstructure (Q-BS) and red-shift phenomenon in ferromagnetic GdN. In the last section~\ref{sec:con}, we conclude with the proposition of a spin resolved ARIPS experiment to be performed in order to validate our results and thus inspire experimental efforts to be carried out in order to study the intriguing temperature dependent properties of GdN.\\

\section{\label{sec:th-mo-cal} THEORETICAL MODEL}

The temperature dependent electronic correlation effects in the conduction bands of GdN are due to the exchange interaction between the electron in extended (5{\it d}) band states and the localized (4{\it f}) electrons which give rise to localized magnetic moments. This situation is covered by the so called ferromagnetic multiband Kondo lattice model (KLM), the Hamiltonian of which reads; \\
\begin{equation}\label{eq:Ham} 
H=H_{kin}+H_{int}
\end{equation}
where \\
\begin{equation}\label{eq:Ho}
H_{kin}=\sum_{ij\alpha\beta\sigma}
T_{ij}^{\alpha\beta}c_{i\alpha\sigma}^{\dagger}c_{j\beta\sigma}
\end{equation}
\indent and
\begin{equation}\label{eq:Hint}
H_{int}=-\frac{J}{2}\sum_{i\alpha\sigma}(z_{\sigma}S_{i}^{z}c_{i\alpha\sigma}^{\dagger}c_{i\alpha\sigma}+S_{i}^{\sigma}c_{i\alpha-\sigma}^{\dagger}c_{i\alpha\sigma})
\end{equation}
\indent The term $H_{kin}$ denotes the kinetic energy of the conduction band electrons, 5{\it d} orbitals in case of GdN, with $c_{i\alpha\sigma}^{\dagger}$ and $c_{i\alpha\sigma}$ being the fermionic creation and annihilation operators, respectively, at lattice site $R_{i}$. The latin letters (i,j,...) symbolize the crystal lattice indices while the band indices are depicted in Greek letters ($\alpha$,$\beta$,..) and the spin is denoted as ${\sigma}(={\uparrow},{\downarrow})$. The multi-band hopping term, $T_{ij}^{\alpha\beta}$ is obtained from an LDA calculation which incorporates in a realistic manner the influences of all those interactions which are not directly accounted for by our model Hamiltonian. It is connected by Fourier transformation to the free Bloch energies $\epsilon^{\alpha\beta}(\bold{k})$
\begin{equation}\label{eq:Hop}
\indent T_{ij}^{\alpha\beta}=\frac{1}{N} \sum_{\bf k}\epsilon^{\alpha\beta}({\bf k}) \hspace{0.2cm} e^{-\rmi{\bf k} \cdot (R_{i}-R_{j})}
\end{equation}
\indent $H_{int}$ is an intra-atomic exchange interaction term being further split into two subterms. The first describes the Ising type interaction between the z-component of the localized and itinerant carrier spins while the other comprises spin exchange processes which are responsible for many of the KLM properties. J is the exchange coupling strength which we assume to be {\bf k}-independent and $S_{i}^{\sigma}$ refers to the localized spin at site $R_{i}$
\begin{equation}\label{eq:Loc-spin}
\indent S_{i}^{\sigma}=S_{i}^{x}+\rmi z_{\sigma}S_{i}^{y} \hspace{0.2cm}; \hspace{0.2cm} z_{\uparrow}=+1, z_{\downarrow}=-1
\end{equation}
\indent The Hamiltonian in eq.~\ref{eq:Ham} provokes a nontrivial many body problem that cannot be solved exactly. The details of the many body analysis along with a model calculation and limiting cases are explained elsewhere (see Ref.~\cite{sharma}). We explain it in brief as follows. \\
\indent Using the equation of motion method for the double-time retarded Green function\cite{zubarev}
\begin{equation}\label{eq:Green-fun}
G_{lm\sigma}^{\mu\nu}(E)=\langle\langle c_{l\mu\sigma};c_{m\nu\sigma}^\dagger\rangle\rangle_{E}
\end{equation}
where l,m and $\mu$,$\nu$ are the lattice and band indices respectively, we obtain higher order Green functions which prevent the direct solution. Approximations must be considered. But, a rather formal solution can be stated as
\begin{equation}\label{eq:Green-matrix}
\widehat{G}_{\bf k}\sigma(E)=\hbar\widehat{I}
[{(E+\rmi 0^{+})\widehat{I}-\widehat{\epsilon}({\bf k})-\widehat{\Sigma}_{\bf k}\sigma(E)}]^{-1}
\end{equation}
where we exclude the band indices by representing the terms in a generalized matrix form on symbolizing a hat over it
\begin{equation}\label{eq:Green-Fourier}
\widehat{G}_{lm\sigma}(E)=\frac{1}{N}\sum_{\bf k}\widehat{G}_{\bf k}\sigma(E) \hspace{0.2cm} e^{-\rmi{\bf k} \cdot (R_{l}-R_{m})}
\end{equation}
\indent The terms in eq.~\ref{eq:Green-matrix} are explained as follows : $\hat{I}$ is an identity matrix and $\hat{\epsilon}({\bf k})$ is a hopping matrix with the diagonal terms of the matrix exemplifying the intra-band hopping and the off-diagonal terms denoting the inter-band hopping. The self energy, $\widehat{\Sigma}_{\bf k} \sigma$(E), containing all the influences of the different interactions being of fundamental
importance, can be understood using site representation :
\begin{equation}\label{eq:self-energy}
\langle\langle[H_{int},c_{l\mu\sigma}]_{-};c_{m\nu\sigma}^{\dagger}
\rangle\rangle=\sum_{p\gamma}\Sigma_{lp\sigma}^{\mu\gamma}(E)G_{pm\sigma}^{\gamma\nu}(E)
\end{equation}
\indent Now, we are left with a problem of finding a multi-band self energy ansatz in order to compute the Green function matrix and thereby calculate the physical quantities of interest like the quasi-particle spectral density (SD)
\begin{equation}\label{eq:SD}
A_{\bf k}\sigma(E)=-\frac{1}{\pi}Im \Tr(\widehat{G}_{\bf k}\sigma(E))
\end{equation}
and the quasi-particle density of states (Q-DOS)
\begin{equation}\label{eq:DOS}
\rho_{\sigma}(E)=\frac{1}{N\hbar}\sum_{\bf k}A_{\bf k}\sigma(E)
\end{equation}
\indent According to our many body theoretical analysis\cite{sharma}, we utilize the multi-band interpolating self energy ansatz (ISA) which is accurately defined in the low carrier density regime, for all coupling strengths and satisfying various limiting cases of the model like the one of ferromagnetically saturated semiconductor as discussed in the Appendix of Ref.~\cite{sharma}. The ansatz is given as :\\
\begin{equation}\label{eq:Mult-SE}
\widehat{\Sigma}_{\sigma}(E)=-\frac{J}{2}M_{\sigma}\widehat{I}+\frac{J^{2}}{4}a_{\sigma}\widehat{G}_{\sigma}(E-\frac{J}{2}M_{\sigma})
\left[\widehat{I}-\frac{J}{2}\widehat{G}_{\sigma}(E-\frac{J}{2}M_{\sigma})\right]^{-1}
\end{equation}
where
\begin{equation}
M_{\sigma}=z_{\sigma} \langle S^{z} \rangle;
\hspace{0.2cm}a_{\sigma}=S(S+1)-M_{\sigma}(M_{\sigma}+1).
\end{equation}
and the bare Green function matrix is defined as :
\begin{equation}\label{eq:Eff-Green}
\widehat{G}_{\sigma}(E)=\frac{1}{N}\sum_{\bf k}\frac{1}{(E+\rmi 0^{+})\widehat{I}-\widehat{\epsilon}({\bf k})} \nonumber\\
\end{equation}
As seen in eq.~\ref{eq:Mult-SE}, we are interested only in the local self energy
\begin{equation}\label{eq:SE-k-ind}
\widehat{\Sigma}_{\sigma}(E)=\frac{1}{N}\sum_{\bf k}\widehat{\Sigma}_{\bf k}\sigma(E)\nonumber\\
\end{equation}
\indent The wave-vector dependence of the self energy is mainly due to the magnon energies $\hbar\omega({\bf k})$ appearing at finite temperature. But, we neglect a direct Heisenberg exchange between the localized spins since we are only interested in the influence of inter-band exchange on the conduction band states in order to study the electronic correlations and not aimed at calculating the magnetic properties via self consistent calculation of localized magnetizations. This can be interpreted as the $\hbar\omega(\bold{k})$ $\rightarrow$ 0 limit. The localized magnetization $\langle S^{z} \rangle$ shall be considered as an external parameter being responsible for the induced temperature dependence of the band states.\\

\section{\label{sec:bndstr-GdN} Bandstructure calculation }

\indent In order to have the single particle excitation energies i.e the hopping matrices, which act as an input in the many body part, and also to find the exchange coupling strength we perform the TB-LMTO\cite{andersen75,andersen84} bandstructure calculations within LDA. GdN crystallizes in a rocksalt structure with an experimental lattice constant of 4.99 $\AA$ and symmetry group Fm-3m. Each $Gd^{3+}$ ion has twelve nearest and six next nearest Gd-neighbours.\\
\begin{figure}[h] 
\centering
\includegraphics[height=4.0cm,width=4.0cm]{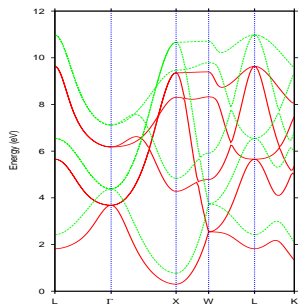}
\caption{\label{fig:lda-bndstr} LDA bandstructure of 5{\it d} conduction bands of GdN. Full (red) lines are spin $\uparrow$ and dotted (green) ones are spin $\downarrow$.} 
\end{figure}
\indent Figure~\ref{fig:lda-bndstr} indicates the calculated spin-dependent 5{\it d} conduction bands of GdN. Our evaluation is restricted only to 5{\it d} bands while considering GdN to be a semiconductor in accordance with Ref.~\cite{hasegawa,lambrecht,ghosh}. But, even after making such  simplifications in our calculation, there are typical difficulties due to the LDA which arise in the treatment of the strongly localized character of the 4{\it f} levels. In order to circumvent this problem we considered the 4{\it f} electrons as core electrons, since our main interest is focused on the response of the conduction bands on the magnetic state of the localized moments. For our purpose, the 4{\it f} levels appear only as localized spins. Moreover, since we are mainly concerned in overall temperature dependent correlation effects, the extreme details of the bandstructure are surely not so important.\\
\vspace*{0.5cm}
\begin{figure}[h]
\begin{center}
\includegraphics[height=4.5cm,width=7.0cm]{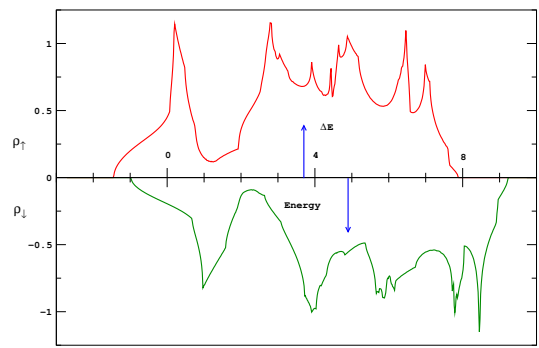}
\caption{\label{fig:exchcoup} Spin-dependent density of states of 5{\it d} bands of GdN. Using the center of gravity of the bands, the exchange splitting amounts to $\Delta$E = 1.237 eV.} 
\end{center}
\end{figure}
\indent In Figure~\ref{fig:exchcoup} the LDA- density of states is displayed. A distinct exchange splitting is visible which can be used to fix the interband exchange coupling constant J in eq.~\ref{eq:Hint}. Assuming that the LDA treatment of the ferromagnetism is quite compatible with the Stoner (mean field) picture, the T=0 splitting amounts to $\Delta$E=JS. On taking the center of gravity of both the bands and alongwith the above assumption we obtain:
\begin{equation}
\Delta E = 1.237 eV \qquad \Rightarrow \qquad J = 0.35 eV 
\end{equation}
\indent In the bandstructure calculations, one cannot switch off the inter-band exchange interaction (eq.~\ref{eq:Hint}) in the LDA code, but one can speculate from the exact T=0 result for empty conduction band\cite{nolting01} that such an interaction leads only to a rigid shift in the $\uparrow$ spectrum while the $\downarrow$ spectrum is remarkably deformed by correlation effects due to spin exchange processes between the delocalized and localized states. So we take from the LDA calculation, which holds by definition for T=0, the $\uparrow$ part as the single-particle input in eq.~\ref{eq:Ho}. Therewith, it is guaranteed that all other interactions which do not explicitly come into our Hamiltonian are implicitly taken into account by the LDA single particle Hamiltonian. On the other hand, a double counting of any decisive interaction is avoided since in the $\uparrow$-spectrum the inter-band exchange (eq.~\ref{eq:Hint}) only shifts the energy zero. Since the methodology used in Ref.\cite{bhattacharjee} remains the same, it is worthwhile to mention that the authors have used paramagnetic bandstructure as an input in their many body part though it is not guaranteed that the paramagnetic input is free of correlation effects.\\
\vspace*{0.5cm}
\begin{figure}[h]
\centering
\includegraphics[height=4.5cm,width=7.0cm]{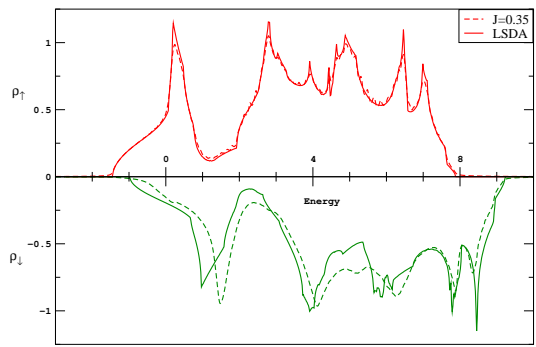}
\caption{\label{fig:jdependos} The same as in Figure~\ref{fig:exchcoup} with additional T=0 results of our combined many body and first principles theory for exchange coupling J = 0.35 eV, shown in broken lines.}
\end{figure}
\indent In order to get a first impression of the correlation effects we have evaluated our theory for T=0 K and the results are presented in Figure~\ref{fig:jdependos}. The $\uparrow$ spectrum is unaffected by the actual value of J and coincides with the respective LDA curve, after a compensation of the unimportant rigid shift ($\frac{JS}{2}$). The slight deviations, seen in the upper part of Figure~\ref{fig:jdependos} are exclusively due to numerical procedures. This shows that our theoretical approach for implementing LDA input into the many body model fulfills the exact limit (T=0, $\sigma$ = $\uparrow$ ) and also circumvents the earlier mentioned double counting problem. The lower half, $\downarrow$ spectrum, of Figure~\ref{fig:jdependos} demonstrates that correlation effects do appear even at T=0 K. Apart from a band narrowing, they provoke strong deformations and shifts with respect to the LDA result. \\

\section{\label{sec:recal-GdN} Model Evaluation}

\indent In this section, we present the numerical results emphasizing the temperature dependent electronic correlation effects in the conduction bands of GdN by combining the first principles TB LMTO bandstructure calculations within LSDA along with the many body theoretical model. \\
\indent The correlation effects are examined as the consequence of a test electron created ( or annihilated ) in an empty (or filled ) band. In our case, we create an electron in an empty conduction band with a consideration of GdN as a ferromagnetic semiconductor. Such effects are studied by calculating the single-electron Green function (eq.~\ref{eq:Green-matrix}). But, apart from electronic sub-system, we also have the magnetic sub-system. The exchange coupling between the itinerant electron and localized spins adds up to the correlation effects as they produce spin-flip transitions and Ising like interactions in addition to the kinetic energy. \\
\indent The numerical calculations remain the same as performed in Ref.~\cite{sharma}. For the sake of brevity, one can understand it as follows. The single particle output obtained from the bandstructure calculations are in the form of Hamiltonian and overlap matrix, posing a generalized eigenvalue problem to be solved. In order to employ such matrices as an input for the many body calculations, one has to perform a decomposition so as to reduce the generalized problem to an eigenvalue problem. Using such an approach, one can factorize the Hamiltonian matrices and then obtain new Hamiltonian matrices using overlap matrices which can be used directly as hopping integrals, $\widehat{\epsilon}({\bf k})$. This factorization and direct use of Hamiltonian matrices circumvent the problem of the somewhat ambiguous decomposition of the conduction band\cite{nolting88,rex,santos} into single non-degenerate subbands which is also used in Ref.\cite{bhattacharjee}. Finally, we use the multi-band self energy ansatz (eq.~\ref{eq:Mult-SE}) to compute the Green function (eq.~\ref{eq:Green-matrix}) and therewith calculate the spectral densities (eq.~\ref{eq:SD}) and densities of states (eq.~\ref{eq:DOS}).\\
\begin{figure}[h]
\centering
\includegraphics[height=4.5cm,width=7.0cm]{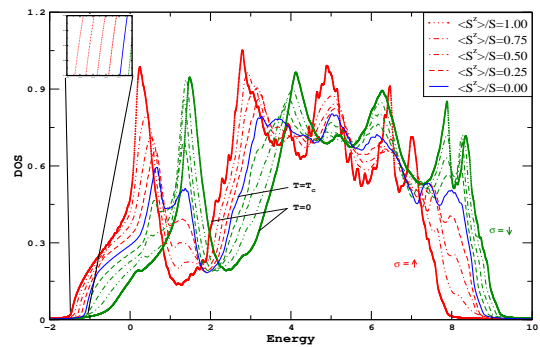}
\caption{\label{fig:qp-dosj35} Quasi-particle density of states (Q-DOS) of the 5{\it d} conduction bands of GdN and as a function of energy for various temperatures (different 4{\it f} magnetizations). The outermost curves belong to T=0 K ($\frac{\langle S^{z} \rangle}{S}$=1).On increasing the temperature they approach each other. The inset shows on an enlarged scale the temperature shift of the lower edge for the case of $\uparrow$. }
\end{figure}
\indent The quasi-particle density of states (Q-DOS) are calculated for J=0.35 eV and  plotted in Figure~\ref{fig:qp-dosj35} for various values of 4$\textit{f}$ magnetizations i.e, different temperature. As observed (from the inset of Figure~\ref{fig:qp-dosj35}), the lower edge of spin - $\uparrow$  Q-DOS performs a shift to lower energies upon cooling from T=$T_{c}$  down to T=0 K which explains the redshift effect of the optical absorption edge for an electronic transition. We find a redshift of 0.34 eV. One should note that a simple mean field treatment with J=0.35 eV and S=3.5 would result in a redshift of 0.6125 eV. Obviously, the  correlation effects drastically reduce this value. A similar redshift phenomenon was discussed by Lambrecht\cite{lambrecht} with a calculated value of 0.30 eV and was also reported experimentally having a value of 0.40 eV by Leuenberger.~\cite{leuenberger} They observed this effect as a shift of the $L_{2}$ edge on a relative comparison between absoprtion spectra measured on a 2000 $\AA$ GdN film at T $>$ $T_{c}$  and T $<$ $T_{c}$ in X-ray magnetic circular dichroism (XMCD) experiment. The $L_{2}$ absorption spectra is sensitive to the polarization of $t_{2g}$ states. In Ref.\cite{bhattacharjee}, the redshift of 0.483 eV was obtained using  J for the lowest band of the bulk GdN and was found to be overestimated.\\
\begin{figure}[h]
\centering
\includegraphics[height=10.0cm,width=8.0cm]{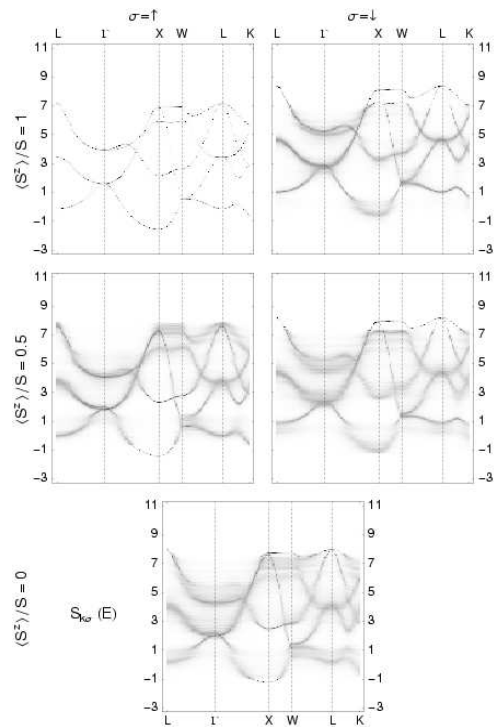}
\caption{\label{fig:qp-bndstrj35} Spin-dependent quasi-particle bandstructure of 5{\it d} conduction bands of GdN for different values of magnetization $\frac{\langle S^{z} \rangle}{S}$.}
\end{figure}
\indent Figure~\ref{fig:qp-bndstrj35} represents the quasi-particle bandstructure for some high-symmetry directions in the first Brillouin zone. The degree of blackening measures the magnitude of the spectral function. As observed in the $\downarrow$ spectrum even at T=0 K ($\frac{\langle S^{z} \rangle}{S}$=1), parts of the dispersions are washed out showing lifetime effects due to correlation in terms of magnon emission and re-absorption with simultaneous spin-flips which can be understood as follows. \\
\indent At T=0 K and empty bands, the addition of a $\uparrow$ electron produces a stable quasi-particle since the $\uparrow$ electron has no chance to flip its spin with the ferromagnetically saturated spin ($\Uparrow$) sub-system. The imaginary part of the self energy vanishes indicating infinite lifetimes. The situation is different for the addition of an $\downarrow$ electron as the $\downarrow$ electron can exchange its spin with the ferromagnetically saturated spin sub-system and have finite lifetime. It can emit a magnon and become a $\uparrow$ electron provided there exist $\uparrow$ electron states which can be occupied after spin-flip process. These are the scattering states which occupy the same energy regions as the $\uparrow$ spectrum. The other possibility for the $\downarrow$ electron to exchange its spin could be by repeated emission and absorption of magnons, thus forming a quasi-particle called magnetic polaron. For strong coupling, the above mentioned elementary processes can even lead to correlation-caused splitting of the dispersion. This is clearly visible at certain parts in Figure~\ref{fig:qp-bndstrj35}.\\
\indent For finite temperatures, the spin sub-system is no longer perfectly aligned. There are magnons in the system that can be absorbed by the itinerant charge carriers provoking quasi-particle damping in the $\uparrow$ spectrum too. The spectral weight gets redistributed due to spin flip term in the exchange interaction with deformations in the density of states. The overall exchange splitting reduces with increasing temperature, until in the limit T
$\rightarrow$ $T_{c}$ ($\langle S^{z} \rangle \rightarrow 0$) , the vanishing 4{\it f} magnetization removes the induced spin asymmetry. \\
\indent In order to have a closer look at the correlations, the spin and {\bf k} -dependent spectral densities are plotted at some high-symmetry points (W,K,X) in the first Brillouin zone and for different temperatures. While the Q-DOS refer to angle averaged photoemission experiment, the ${\bf k}$-dependent spectral densities refer to the angle resolved part.\\
\vspace*{0.3cm}
\begin{figure}[h]
\centering
\includegraphics[height=4.5cm,width=7.0cm]{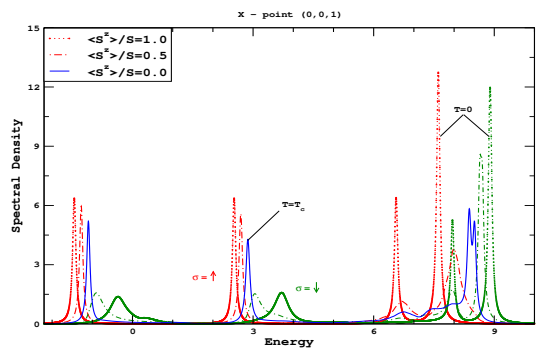}
\caption{\label{fig:x-pt} Spectral density at X - point. The case of $\frac{\langle S^{z} \rangle}{S}$ = 1.0 is shown by dotted lines, while $\frac{\langle S^{z} \rangle}{S}$ = 0.50 is denoted as broken lines and the full lines represent $\frac{\langle S^{z} \rangle}{S}$ = 0.0 }
\end{figure}
\indent As seen in Figure~\ref{fig:x-pt}, well defined quasi-particle peaks appear with an additional spin split below
$T_{c}$.  In accordance with the quasi-particle bandstructure, it is observed in Figure~\ref{fig:x-pt} that four sharp peaks appear at T=0 K for $\uparrow$ spectrum and though already strongly damped (at lower energies), the same peak sequence comes out in the $\downarrow$ spectrum. The exchange splitting collapses for T $\rightarrow$ $T_{c}$ and the quasi-particle damping increases with increasing temperature. For T=$T_{c}$ (full lines), the spectrum is washed out at higher energies exhibiting strong correlation effects.\\
\indent Similar explanations hold for the other high-symmetry points (Figure~\ref{fig:w-pt} and Figure~\ref{fig:k-pt}). There's a strong damping as seen along the entire energy spectrum. And some of the peaks (lower energy peaks in Figure~\ref{fig:w-pt}) are so strongly damped that they may not be visible in the inverse photoemission experiment. Altogether, the 5d spectral densities exhibit drastic temperature dependencies, what concerns the positions and the widths of quasi-particle peaks.\\
\begin{figure}[h]
\centering
\includegraphics[height=4.5cm,width=7.0cm]{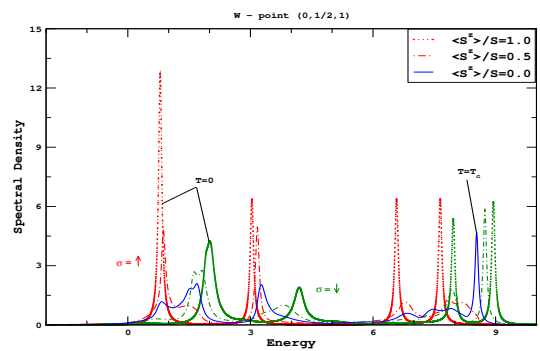}
\caption{\label{fig:w-pt} The same as in Figure~\ref{fig:x-pt} but for W - point.}
\end{figure}
\vspace{0.2cm}
\begin{figure}[h]
\centering
\includegraphics[height=4.5cm,width=7.0cm]{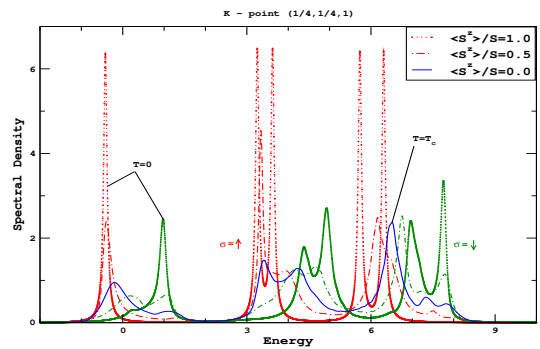}
\caption{\label{fig:k-pt} The same as in Figure~\ref{fig:x-pt} but for K - point.}
\end{figure}
\vspace{0.2cm}

\section{\label{sec:con} SUMMARY and CONCLUSION}

\indent We investigated the temperature dependent electronic correlation effects on the 5{\it d} conduction bands of GdN. In this respect, we started with multi-band Kondo lattice Hamiltonian which describes an exchange interaction between the localized moments and itinerant conduction electrons. Using the equation of motion approach in terms of Green function, we tried to evaluate the multi-band KLM Hamiltonian. In order to find the solution, i.e, Green function, there was a need to find a self energy ansatz which incorporated all the interactions of the system. We used an Interpolating Self energy Ansatz (ISA)\cite{nolting01} for the multi-band\cite{sharma} case which was found to be reliable for low carrier densities and all coupling strengths and which fulfilled important limiting cases of the model. \\
\indent We further performed bandstructure calculations using TB LMTO program in order to obtain hopping matrices which served as an input in the many body analysis. Though there has been a wide discrepancy regarding the nature of the electronic ground state of GdN, we considered it to be a semiconductor and thereby concentrating only on the empty 5{\it d} bands and the correlation effects due to their interaction with 4{\it f} localized moments. From the bandstructure calculations, we also obtained the inter-band exchange coupling strength as a consequence of exchange based splitting of the 5{\it d} conduction bands.\\
\indent The first principles band structure calculations were combined with the many body theory in order to calculate the Green function and therewith some of the physical properties of interest in GdN like quasi-particle spectral densities (Q-SD) and densities of states (Q-DOS) are estimated for different values of magnetization and exchange coupling strength. A redhift of 0.34 eV for exchange coupling J=0.35 eV was calculated and found to be in close comparison with other theoretical prediction and experimental value reported in the literature. Strong temperature dependent correlation effects were observed over the whole energy spectrum. Our results can be compared with spin-dependent ARIPS of the conduction bands of GdN and thus we would like to instigate experimental efforts in order to study this interesting compound. \\

\ack{We would like to thank Prof. P.A.Dowben (UNL, USA), Prof. W.R.L Lambrecht (CWRU, USA) and PD Dr. V.Eyert (Universit$\ddot{a}$t Augsburg, Germany) for helpful discussions and comments on the electronic bandstructure of GdN.}

\end{document}